\documentclass[prl%
reprint, 
twocolumn, 
secnumarabic,
tightenlines,
amssymb,
amsmath,
nobibnotes,
nofootinbib, 
floatfix,
aps, 
showpacs,showkeys,
]{revtex4-2}
\usepackage{hyperref}
\usepackage{multirow}%
\expandafter \ifx \csname package@font\endcsname\relax\else
\expandafter \expandafter \expandafter
\usepackage
\expandafter \expandafter
\expandafter{\csname package@font\endcsname}%
\fi
\usepackage[normalem]{ulem} 
\usepackage{textcomp}
\usepackage{gensymb}

\usepackage{forest}

\hypersetup{
	colorlinks=true,
	linkcolor=blue,
	citecolor=blue,
	urlcolor=blue,
	pdfborder={0 0 0}
}

 \usepackage{hyperref} %
\usepackage{cleveref}
\usepackage{graphicx} %

\begin{abstract}
Transfer learning (TL) allows a deep neural network (DNN) trained on one type of data to be adapted for new problems with limited information. We propose to use the TL technique in physics. The DNN learns the  details of one process, and after fine-tuning, it makes predictions for related processes. We consider the DNNs, trained on inclusive electron-carbon scattering data, and show that after fine-tuning, they accurately predict cross sections for electron interactions with nuclear targets ranging from  helium-3 to iron.  
\end{abstract}

\begin{document}

\title{Electron-nucleus cross sections from transfer learning}

\author{Krzysztof M. Graczyk}
\email{krzysztof.graczyk@uwr.edu.pl}

\author{Beata E. Kowal}

\author{\\Artur M. Ankowski}
\author{Rwik Dharmapal Banerjee}
\author{Jose Luis Bonilla}
\author{Hemant Prasad}
\author{Jan T. Sobczyk}

\affiliation{Institute of Theoretical Physics, University of Wroc\l aw, plac Maxa Borna 9,
50-204, Wroc\l aw, Poland}

\date{\today}%

\maketitle

Transfer learning (TL) is a widely studied phenomenon in psychology and education~\cite{STEINER200115845}. It refers to the ability of a person who has learned skills in one specific field to easily acquire skills needed in related areas of life. The concept of TL has been recognized in the artificial neural network community since 1976~\cite{Bozinovski2020}. Currently, TL plays a significant role in the development of modern deep neural network (DNN) systems~\cite{zhuang2020comprehensive}. There are numerous successful applications of TL in deep learning~\cite{zhuang2020comprehensive,Weiss2016}. One example could be DNN, trained on non-medical data, which, after fine-tuning, is used to analyze medical photos and diagnose cancer. 

This letter  demonstrates that the TL technique can be efficient in modeling physical systems. The information acquired by a network that has learned the physics of a specific phenomenon  is transferred, using deep learning methods, to describe related  problems. 

We focus on the problem of modeling electron-nucleus scattering. Our motivation comes from studies of neutrinos' fundamental properties. The next generation of neutrino-oscillation experiments, DUNE~\cite{DUNE:2020lwj} and Hyper-Kamiokande~\cite{Hyper-KamiokandeProto-:2015xww}, aim to determine the value of the Dirac~phase and shed light on the question of the dominance of matter over anti-matter in the Universe. Their success is contingent upon the accuracy of our estimates of neutrino cross sections for interactions with nuclear targets, which is required to translate the observed neutrino-antineutrino differences between the appearance-event rates to the oscillation probabilities. In particular, interactions of (anti)neutrinos with argon will be detected in DUNE, and (anti)neutrino-oxygen scattering in water is relevant for Hyper-Kamiokande.

A number of factors contribute to the complexity of the problem of neutrino interactions with nuclei in the few-GeV energy region, relevant for the oscillation studies~\cite{Ankowski:2020qbe}. Since neutrino-nucleus cross sections are not known with sufficient precision, modeling nuclear effects remains difficult. The measured neutrino scattering cross sections are averaged over broad fluxes, which greatly diminishes differences between  various interaction channels at the inclusive level. Different interaction mechanisms yield the same final states, making a piecewise model development approach unsuitable. Moreover, the process of scattering involves multicomponent vector and axial currents, which are impossible to disentangle and determine based on neutrino data alone.

Recently, a  consensus has emerged to make full use of the similarities of electron and neutrino interactions at the nuclear level~\cite{Ankowski:2022thw}, to greatly reduce the uncertainties of neutrino cross sections originating from nuclear effects, such as the description of the ground state  or  final-state interactions.

The scarcity of the electron-scattering measurements for argon and oxygen is currently a major obstacle~\cite{Dai:2018xhi,O'Connell:1987ag,Anghinolfi:1996vm}. On the other hand, a large body of data is already available for carbon~\cite{RevModPhys.80.189}. In this paper, we employ the concept of the TL to demonstrate that DNNs trained on carbon data can accurately predict the cross sections for nuclear targets ranging from helium-3 to iron, provided that even scarcely distributed measurements span the relevant kinematic region,  and that they are sufficient to demonstrate that the same mechanisms of interaction apply to the considered targets and to carbon. Although the problem we consider is specific, the proposed method is general and can be applied to other particle and nuclear physics studies. 

Before we delve into the electron-nucleus scattering, let us first clarify the concept of the TL. A DNN  transforms an input vector (or tensor) into an output (scalar, vector, or tensor). It can be represented by a graph of connected neurons grouped in multiple layers. A single neuron transforms the input into a scalar. 
\begin{figure*}
\includegraphics[width=\textwidth]{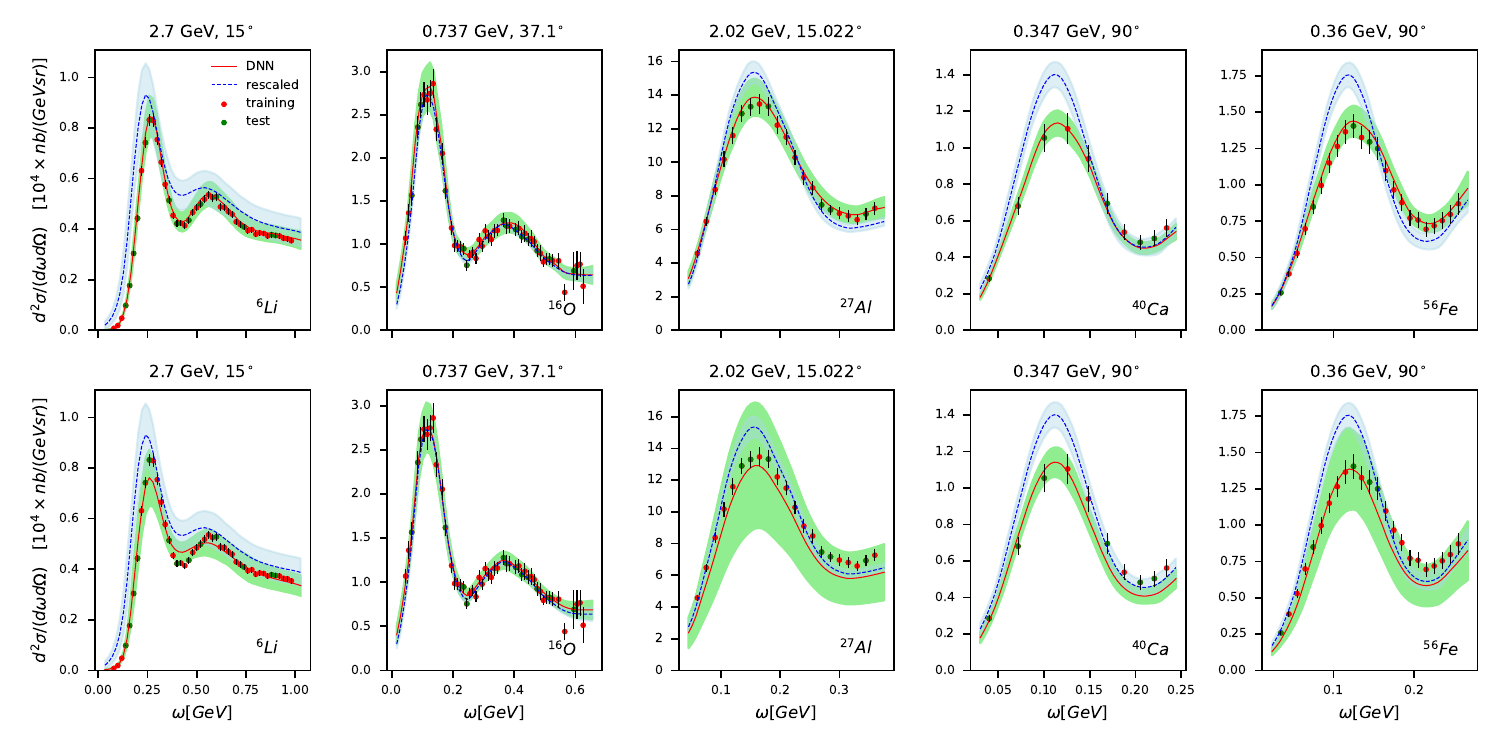}
\caption{Double-differential cross section ${d^2 \sigma}/{d\omega d\Omega}$ for inclusive electron scattering on lithium~\cite{Heimlich:1973}, oxygen~\cite{O'Connell:1987ag}, aluminum~\cite{Day:1993md}, calcium~\cite{Williamson:1997}, and iron~\cite{Meziani:1984is} for selected kinematics. The fits obtained for the proportion of training to test datasets 7:3. The red line denotes the DNNs predictions, and the green area denotes $1\sigma$ uncertainty. The results for  DNN with all/two last layers fine-tuned are shown in the top/bottom row. The electron-carbon DNN predictions multiplied by factor $A/12$ are shown by blue-dashed lines corresponding to $1\sigma$ uncertainty denoted by a light blue area. The red (green) points represent the training (test) dataset. The $^6$Li and $^{27}$Al data were collected at kinematic setups of great importance for DUNE, and the $^{16}$O, $^{40}$Ca, and $^{56}$Fe data probe the kinematic region relevant for the Hyper-Kamiokande and T2K experiments.}
\label{Fig:1}
\end{figure*}

One of the most intriguing general properties of deep learning is that neurons in the first layers of DNN extract basic features from the input, while the units in the last layers combine these features into more complex ones. This mechanism is known as representation learning~\cite{representation_learning_review}. For example, early layers of convolutional neural networks, used in picture analysis, detect edges and corners, while deeper layers combine them to recognize abstract features.

Assuming that the ability to extract basic features is transferable across a broad range of problems, a network trained on one data type can be fine-tuned to analyze other related data.  
A common approach involves replacing the last layers of a pre-trained network (on given informative dataset) with a new one and training them with data from the new problem~\cite{tan2018survey}. Another possibility is to fine-tune all parameters of the pre-trained network slightly.
These approaches offer several benefits, the most significant being the ability to train a new model using a small dataset, even if it only partially covers the domain of interest. TL holds significant potential for enabling new applications across diverse domains.

To show that TL works effectively and can be utilized to model lepton-nucleus scattering, we consider the DNNs trained on the inclusive electron-carbon cross section data~\cite{Kowal:2023dcq}. The obtained model predicts cross sections in the kinematic range covering quasielastic peak, resonance region, and the onset of deep inelastic scattering.  Study in this paper is based on the bootstrap model~\cite{Kowal:2023dcq}, which showed very good extrapolation abilities. In this approach, we collected an ensemble of $50$ networks. Each of them consists of ten layers of hidden units. The mean and variance over all network predictions in the ensemble give the model's predictions and predictive uncertainty, respectively.
\begin{figure*}
\includegraphics[width=\textwidth]{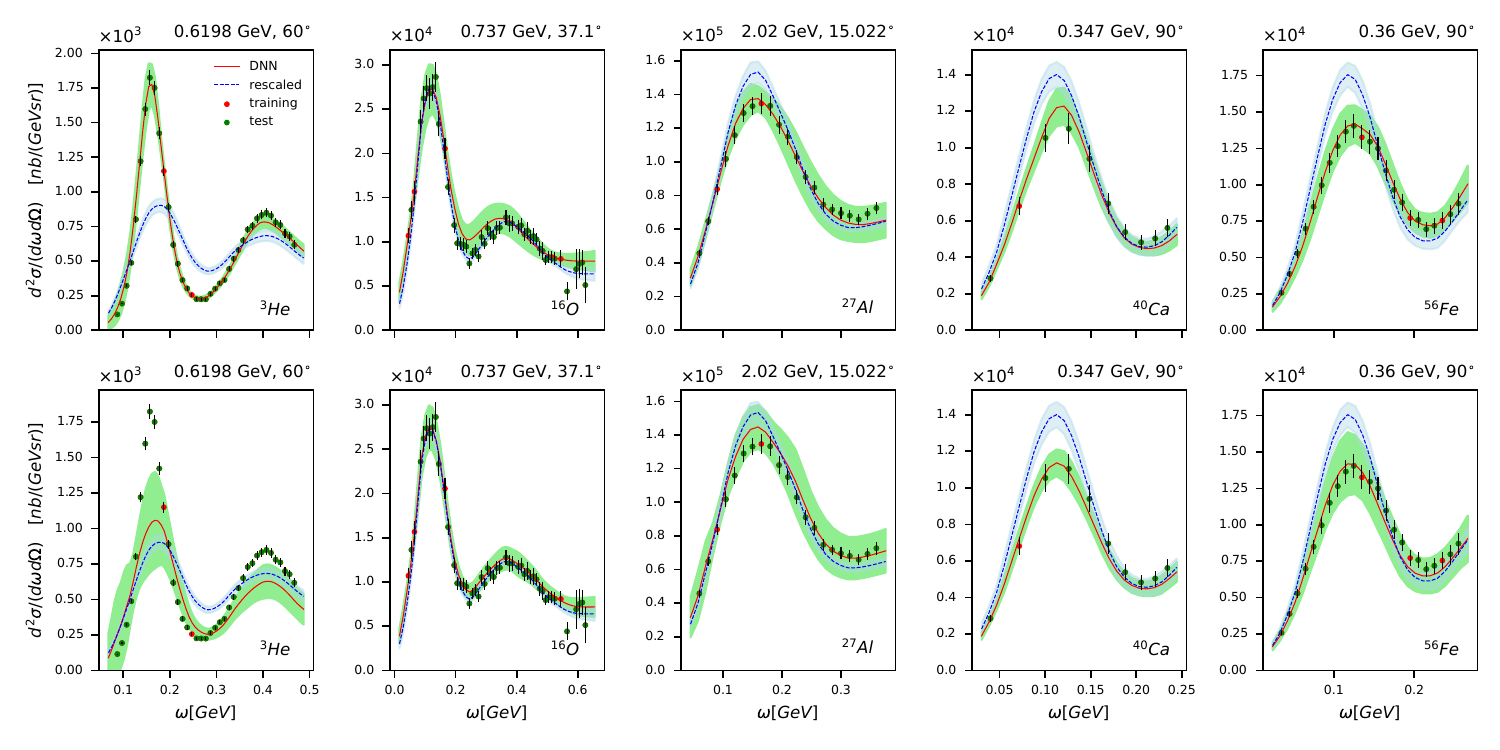}
\caption{Same as in Fig. \ref{Fig:1} but  for the proportion of training to test datasets 1:9 and for $^{3}$He~\cite{Marchand:1985us} instead of $^{6}$Li}.
\label{Fig:2}
\end{figure*}

We hypothesize that the network, trained on electron-carbon data, has learned nuclear structure features that are broadly applicable to other medium-mass nuclei. While these nuclei share common reaction mechanisms—such as quasielastic scattering, resonance excitation, and deep inelastic scattering—they also exhibit important structural differences that affect the electron-scattering cross section. Our study includes both symmetric and asymmetric nuclei, examples of open- and closed-shell configurations, as well as targets with varying average separation energies and momentum distributions.

These differences are likely to be accumulated and captured in the final layers of the network rather than in the initial layers. If the DNN encodes substantial information about nuclear physics, it may fill the gaps in our experimental knowledge, until measurements are performed. Although DNNs can serve as sophisticated interpolation and extrapolation tools, one needs to keep in mind that new observations remain crucial for validating their predictions and discovering new physical phenomena.

Indeed, our model predictions, being solely data-driven, require careful examination. This involves comparing the predictions to unseen measurements, evaluating the fit quality using the $\chi^2$ statistic, and estimating predictive uncertainty. Additionally, we assess the impact of TL through layer-wise fine-tuning.

The following framework is applied to investigate how TL works for electron-nucleus scattering: The pre-trained DNNs are retrained using new measurements in two distinct scenarios. Firstly, the parameters in all layers of the DNNs are optimized (full fine-tuning). In the second scenario (shallow fine-tuning), we only optimize the parameters in the last two layers (fine-tuning only the last layer was ineffective). To train the DNN, we divide randomly the data into training and test sets. Initially, the split is 7:3, but to gauge the effectiveness of a limited training dataset, we also consider a split 1:9.

We analyze measurements of the inclusive cross section for electron scattering off helium ($^{3}$He), lithium ($^6$Li), oxygen ($^{16}$O), calcium ($^{40}$Ca), aluminum ($^{27}$Al), and iron ($^{56}$Fe) for energies up to $8$~GeV. The data are taken from the database \cite{data_repository} and they are in the form of double differential cross sections $d^2\sigma /d \omega d \Omega $ depending on incoming electron energy ($E$), energy transfer ($\omega$), and scattering angle ($\theta$).
We exclude low-energy data points from each target dataset as explained in~\cite{Kowal:2023dcq}. 
The electron-carbon scattering data span a broad kinematic range that largely overlaps with data for other targets but misses some, such as aluminum at $\theta \approx 53^\degree$.
For each target, we consider a separate ensemble of  DNNs that are fine-tuned for each process to describe new measurements. To adapt the pre-trained (on the electron-carbon scattering data) DNN models to analyze the other target data, we multiply the pre-trained DNN output by the scale factor of $A/12$, where $A$ corresponds to the  mass number of a given target, 12 is the  mass number of carbon. The DNN takes for the input $E$, $\omega$, $\theta$, and two  dependent  variables, $\cos\theta$ and four-momentum transfer squared.

To fine-tune the model, we  ran the optimizer only for several hundred epochs, ranging from 450 to 1,200, compared to over 40,000 epochs required for optimizing the pre-trained model. We consider the same statistical framework as in Ref.~\cite{Kowal:2023dcq}. The loss is given by the $\chi^2$, including statistical and point-to-point systematic uncertainties. We include a penalty term for each independent dataset that allows us to control the relative normalization parameters and account for systematic normalization uncertainties. For details see Supplemental Material~\cite{supplemental}.

We first examine the results for the lithium, oxygen, aluminum, calcium, and iron targets, using a 7:3 training-to-test dataset split. We observe a perfect agreement between the predictions of the fine-tuned DNNs  and the data in this case, as shown in Fig.~\ref{Fig:1} and in supplement~\cite{supplemental}. The best agreement is achieved  for full fine-tuned models, but the shallow fine-tuned models work well, too.

In all figures, we plot the pre-trained DNN model predictions multiplied by the factor of $A/12$ to illustrate how the fine-tuning procedure changes the pre-trained model predictions. There are well-known similarities between the $^{16}$O and $^{12}$C nuclei. Therefore, in this case, simple re-scaling of the cross sections produces a reasonable agreement. For the other targets, it is necessary to fine-tune the model  significantly. In fact, our DNN models predict the cross sections in quasielastic peak, dip, resonance, and the onset of deep inelastic regions. Therefore, a simple scaling cannot be applied.

The power of TL is revealed once the information in the training dataset is limited. To illustrate this, we consider a 1:9 split between the training and test datasets, see Fig.~\ref{Fig:2} and supplement~\cite{supplemental}.
The full fine-tuned model can reproduce the cross~section's shape and strength with noteworthy accuracy for almost all targets, even when there are few or no training points in a particular set of measurements. For the shallow fine-tuning the DNN predictions also agree with the data with reasonable accuracy. 

The fact that the full fine-tuned model works better shows that the distribution of the nuclear structure information in the network is nontrivial. Both fine-tuning scenario models work well for oxygen, calcium, and iron. For lithium, and aluminum, the full fine-tuning leads to the model significantly more accurate than the shallow fine-tuned one. 

To test the limitations of the method, we consider the $^3$He target, which exhibits strong few-body correlations and short-range dynamics that are more pronounced than in heavier nuclei, making it significantly different from $^{12}$C. Although helium-3 differs markedly from carbon, TL method performs well when the model is fully fine-tuned. However, in the case of shallow fine-tuning, the method fails to reproduce the data accurately. This indicates that the structural differences between $^3$He and $^{12}$C are substantial enough to require a deeper re-optimization of the model.

The success of the TL-based approach has a twofold origin.  Firstly,  DNN apparently can identify relationships in electron scattering data~\cite{RevModPhys.80.189}, 
for instance, those described by the superscaling hypothesis~\cite{Donnelly:1999sw}. The DNN can recognize these relationships by analyzing electron-carbon data and using minimal input information from measurements of other targets. Secondly, DNNs are known for their strong generalization abilities~\cite{zhang2017understanding}, although the exact mechanism behind this still needs to be well understood. This property is related to the fact that DNNs have a much larger number of parameters than the data constraints. Therefore, when the model is trained, its parameters are only partially constrained by the data, leading each parameter to contain a fraction of random noise, which prevents overfitting and produces a good model generalization ability.

Summarizing, our study demonstrates that deep neural networks can effectively learn the physics of electron-carbon scattering. The network encodes fundamental properties of the nuclear medium, which can be leveraged through the TL techniques to model related processes, even with limited information. This method is rooted in the fundamental property of deep learning: DNNs can extract basic features from data and use them to build more complex abstract features. On the other hand, our method relies on similarities between reference and other nuclear targets; thus, significant structural differences between nuclei may reduce its efficiency.

Our paper focuses on a specific aspect of lepton-nucleus scattering physics, but the potential of the TL extends far beyond this application. It can be effectively utilized in various other domains within nuclear and particle physics and other scientific fields. In particular, our proposed technique holds great promise in accurately modeling neutrino-nucleus scattering. There are significant parallels between the interactions of electrons and neutrinos with nuclei, suggesting that a model trained on electron-nucleus scattering data can be leveraged, through the TL, to model neutrino-nucleus cross sections.  We plan to explore this issue in detail in our future publications. Such a novel methodology has the potential to significantly enhance our understanding of neutrino interactions with nuclei, which is crucial for next-generation neutrino long-baseline oscillation experiments.

\textit{Acknowledgments}---This research is partly (K.M.G., A.M.A., J.T.S.) or fully (B.E.K., R.D.B., J.L.B, H.P.) supported by the Na{\-}tional Science Centre under grant UMO-2021/41/B/ST2/ 02778. K.M.G is partly supported by the ``Excellence Initiative – Research University" for the years 2020-2026 at the University of Wroc\l aw.

\normalem
\begingroup
\sloppy
\setlength{\emergencystretch}{1em}

\end{document}